# New vision of convection induced freckle formation theory in Nickel-based superalloys by electron microscopy

Shuai Wang [a, †], Yuliang Jia [b, †], Yongzhe Wang [c, †], Yongjia Zhang [d], Lan Ma [e], Feng Cheng [a], Yi Zeng [c,*], Xu Shen [d], Yingliu Du [b], Binghui Ge [a,*]

[a] Information Materials and Intelligent Sensing Laboratory of Anhui Province, Institutes of Physical Science and Information Technology, Anhui University, Hefei, 230601, China

[b] Anhui Yingliu Hangyuan Power Technology Co., Ltd, Huoshan, 237200, China

[c] The State Key Lab of High Performance Ceramics and Superfine Microstructure, Shanghai Institute of Ceramics, Chinese Academy of Sciences, 1295 Dingxu Road, Shanghai 200050, China

[d] State Key Laboratory of Materials Processing and Die & Mould Technology, School of Materials Science and Engineering, Huazhong University of Science and Technology, Wuhan, 430074, China

[e] Oxford Instruments Technology China

[†]Equally contributed

* The author to whom correspondence should be addressed: bhge@ahu.edu.cn (Binghui Ge); zengyi@mail.sic.ac.cn   (Y. Zeng)

# Abstract

Freckles, one of the common defects in blades used in heavy duty gas turbines, hugely deteriorates blades' mechanical properties and liability under service conditions. Thermal-solutal convection theory is a widely adopted formation mechanism but few solid experimental evidences were reported. Here for the first time we systematically studied the microstructure of 117 grains in freckle chains from four different Nickel-based superalloys of either single crystal or directionally solidified alloys. The relationship between the internal stress and the misorientation throughout the freckle chains is studied by means of state-of-the-art electron microscopy. All results give new experimental proof to the theory of thermal-solutal convection, which is further supported by the fact that borides at the boundary are randomly orientated to alloys. Our results enrich the methodology of freckle study, providing a new sight of the formation mechanism of casting defects.

# 1. Introduction

Working under high temperatures and the complex stress-corrosive environments, turbine blades put forward a harsh requirement on the material. Nickel-based superalloys are ideal for these applications thanks to their extraordinary properties such as high temperature strength, creep and fatigue properties [1-3]. However, a series of casting defects formed during the directional solidification seriously affect the service performance [4, 5]. For example, as to the heavy duty gas turbine (HDGT) blades, freckles are one of the common casting defects, that is, equiaxed crystal chains with random orientation grow on the blade surface along the solidification direction, which introduces transverse grain boundaries to alloys [6, 7].

How to predict or decrease the formation of freckles attracts wide interest. Two factors are found to increase the tendency of freckle formation: first, to achieve higher thermal efficiency of the gas turbine, the size of the turbine blade increases, and the structure of the aero-engine is designed to be more complicated; second, more refractory elements such as W and Re are required in recent years to improve the endurance temperature of the blade [2, 7, 8].

Theoretically, defined as the ratio of driving force to resistance in solute convection, Rayleigh number is considered as the critical value to predict the formation of freckle [9]. For different alloy systems, however, the Rayleigh number should be corrected according to solidification conditions as an important reference [10-12]. Experimentally, researchers tried different methods to avoid freckles, such as inversion solidification, tilting the mold and adding a magnetic field, etc. [13-15], but most of them failed to achieve ideal results. Therefore, deeply understanding the formation mechanism of freckles is of great significance.

Thermal-solutal convection theory is mostly accepted to explain the formation of freckles [2]: solute segregation causes density inversion and solute convection in the channel, leading to the remelt or even fracture of dendrites, which eventually accounts for the presence of freckles [16]. Therefore, the formation of freckles requires two prerequisites, an open solute channel and sufficient thermal-solutal convection. It is reported that the overgrowth, deflection and remelting of dendrites result in open solute channels, whose survival depends on the competition among solute transport, lateral heat flux and the growth of dendrites [17-19]. On the other hand, considering that the density inversion is the basis of solute convection, the partition coefficient, the ratio of mass fraction of the element between dendrite and inter-dendrite, is an important parameter to evaluate the effect of different elements on freckle initiation [7, 20]. In the experiments, it was found that the addition of element C, W, Re and Hf notably affect the extent of solute segregation and change the convection intensity, thus regulating the formation of freckles [8, 21-23].

In essence, freckles are believed to be the result of fracture rotation of secondary dendrites. During the directional solidification, secondary dendrite forms and is then put in a shear deformation under coupling effect from both remelt and solute convection [24, 25]. As to remelting, solute (such as Al and Ti) enriches at the root of secondary dendrite during solidification, which changes the local composition balance of the solid-liquid interface, then the root remelts [26-28]. While for the stress induced by convection, it seems easier to understand that the thermal-solutal convection in supercooled melts makes the secondary dendrite subjected to shear deformation and causes the secondary dendrite to bend [29-32]. With the help of the remelting, the critical load value can be easily reached, thus the secondary dendrites can be fractured and freckles are formed [33, 34]. As to the above-mentioned solute-convection theory, some experimental and computational results were reported on macroscopic or mesoscopic scale [7, 18, 33]. Nevertheless, the formation of freckles is a complex process, and most of the researches can only focus on one certain snapshot of the whole freckle formation process, and especially there is still a lack of sufficient knowledge about its mechanism from the point of view of microstructure.

Electron microscopy is an important method to characterize the microstructure of materials. It can give not only the information of composition and the partition of different elements but also the information of local crystal structure, including the orientation and strain distribution. Then the formation mechanism of freckles can be deeply understood from multiple perspectives. Considering that dendrites will deform under the solute scouring in solute convection theory, and fragments may rotate freely in the mushy zone after fracture [35], it should be believed that the strain in these grains and the misorientation of freckle grains relative to the matrix should have a specific distribution. Therefore, in this paper, we will use scanning electron microscope (SEM), energy dispersive x-ray spectroscopy (EDS), electron backscattering diffraction (EBSD), and aberration-corrected electron microscope to analyze the misorientation angle (MA) and strain in directionally solidified alloys, 247LC-DS, GTD111-DS, and single crystal alloys, RenéN4 and CMSX-4. In addition, considering that borides in alloys usually intergrow with a specific orientation relationship to the matrix [36-38], borides at the grain boundary between freckles and matrix can also be used as a marker to further understand the movement of freckle grains and then their formation mechanism. Hereafter, unless otherwise specified, all samples refer to directionally solidified alloy 247LC-DS.

## 2. Experimental methods

2.1 Preparation of the blades

Four Nickel-based superalloys studied in this paper are 247LC-DS, GTD111-DS, RenéN4 and CMSX-4, among which 247LC-DS and GTD111-DS are directionally solidified alloys, while RenéN4 and CMSX-4 are single crystal alloys (see the nominal chemical compositions in Supplementary Table S1). Blades were prepared by Bridgman directional solidification technology, in which single crystal blades were obtained by the spiral crystal selection method, and the drawing speed was 4 mm/min. After solidification, shelling and sandblasting were carried out.

2.2 Sample preparation and data acquisition

The surface of the blade was etched with 50% HCL+50% $H_2O_2$ solution to show the freckles. The freckle chains in the blade tenons and their surrounding matrix were cut into block samples in the size of $10 \times 10 \times 3$ mm by an electric spark cutter. Samples were polished step by step with silicon carbide sandpaper from 240 to 5000 #, followed by 1 μm diamond grinding paste until the surface was scratch-free. Then, they were etched in 20 g $CuSO_4$ + 100 ml HCL + 80 ml $H_2O$ for 20s for optical microstructure observation on a Nikon ECLISE E200.

EBSD samples were prepared by mechanical polishing followed by electropolishing. The mechanical polishing procedure was the same as above. Electropolishing was conducted at 25 V, -25°C for 20s in the electrolyte of 10% $HCLO_4$ + 90% $C_2H_5OH$. EBSD test was carried out on Carl Zeiss Crossbeam 550L equipped with Oxford Symmetry S2 EBSD under the acquisition condition of 20 kV and 25 nA in beam current, all EBSD tests used a step size of 1.5 μm. After the acquisition, the post-processing analysis of EBSD data was carried out with AZtec Crystal. To analyze/quantify the internal stress along the freckles, HR-EBSD was performed using a mega-pixel pattern size of $1024 \times 1344$ in a speed of 20 Hz. The internal stress was then calculated using a Digital Image Correlation (DIC) method applied on the patterns. More detailed algorithms could be referred to the previous article [39, 40]. Ultim Max EDS of Oxford Instrument was used to collect at 20 kV, 2 nA for the purpose of compositional distribution in the samples/investigating element segregation at the dendrites and inter-dendrites.

Lamellas for transmission Kikuchi diffraction (TKD) and transmission electron microscope (TEM) were prepared through Carl Zeiss Crossbeam 550L focus ion beam (FIB). TKD was analyzed under 30 kV with a beam current of 30 nA. Bright field (BF) images, aberration-corrected high angle annular dark field (HAADF) images and electron energy loss spectrum (EELS) were obtained on Titan Themis Z equipped with double aberration corrector and the operating voltage is 300 kV. When Z-contrast images at the atomic scale are taken, the convergence angle of the electron beam is 25 mrad.

## 3. Results

3.1 Morphology and composition segregation of freckles

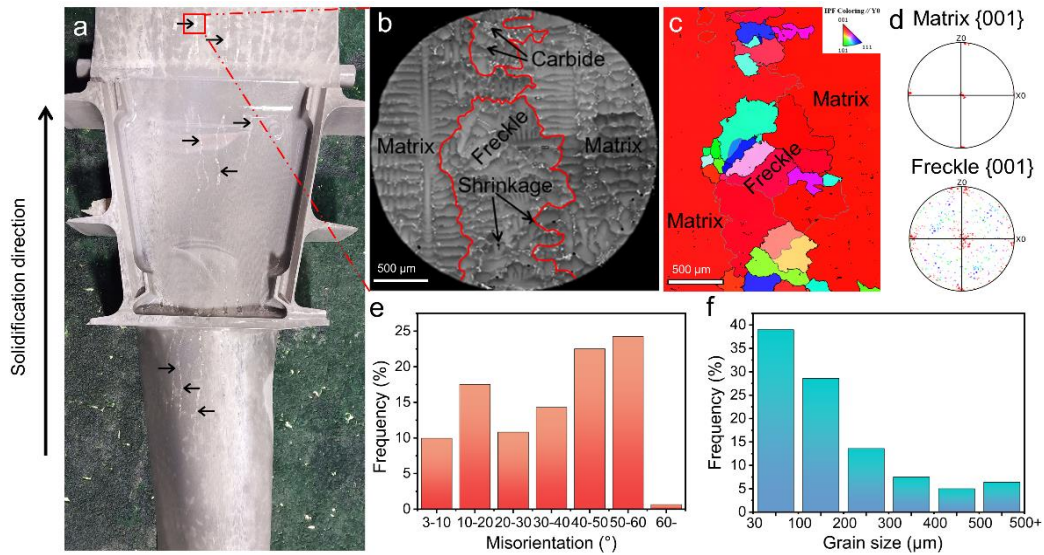

**Fig. 1.** Dendrite morphology and grain orientation distribution of freckle chain (the direction of solidification is from bottom to top).

(a) Photograph of the turbine blade casting illustrating the presence of freckle chains, and freckle chains marked with black arrows. (b)-(c) BSE and orientation image of freckle chain and the matrix corresponding to the area outlined by a red square in (a). The interface is outlined by red curves, and carbides and shrinkage cavities are marked with black arrows. Color legend of orientation is inset on the top right. (d) Pole figure of matrix and freckle, respectively, corresponding to the area in (b). (e)-(f) Misorientation of freckle grains relative to the matrix and freckle grain size in four brands of samples.

Fig. 1a shows a 247LC-DS blade, and the areas like chains with the white contrast along the solidification direction (vertical direction in the figure) on the blade surface are freckle chains, indicated by black arrows. The morphology of dendrites can be observed by SEM and optical microscope as shown in Fig. 1b and Supplementary Fig. S1. The dendrites of the matrix are arranged neatly at the two sides, while the middle area, like a channel, contains carbides (B-containing alloy contains borides), shrinkage cavities and randomly distributed grains, i.e. freckles.

Fig. 1c shows the crystal orientation distribution from Y0 (IPFY) direction, parallel to the solidification direction. Different color in this figure indicates different crystallographic orientation, so a freckle is composed of grains with random orientation, in opposition to the single-crystallized matrix. The same conclusion can be obtained from Fig. 1d, the pole figure (PF) of Fig. 1c, and the other three alloys show similar results. The MA relative to the matrix of all freckle grains in four kinds of alloys was counted, and the statistical results are shown in Fig. 1e. Two-thirds of freckle grains have an angle to the matrix in the range of 30-60 degrees. According to the statistics of freckle grain size with the equivalent circle diameter as a reference [41], as shown in Fig. 1f, the size of freckle grains is mostly within the range of 30-200 μm.

As we all know, Nickel-based superalloys usually contain more than ten kinds of alloy elements, and different elements have different distributions in the alloy (element mapping shown in Fig. S2a), which has a significant impact on the properties of the alloy, such as the famous Re effect [42, 43]. The segregation degree of an element in the alloy between dendrite and inter-dendrite can be expressed by the element segregation coefficient $K_i$, $K_i = C^i_{dendrite}/C^i_{inter\text{-}dendrite}$, where $C^i_{dendrite}$ and $C^i_{inter\text{-}dendrite}$ are the mass fraction of the element between dendrite and inter-dendrite, respectively. If $K_i > 1$, the elements segregate in dendrites; if $K_i < 1$, the elements segregate in inter-dendrites. EDS data were collected at dendrites and inter-dendrites in the matrix and freckle areas, respectively, and the segregation coefficient of each element in the matrix and freckles for four alloys were calculated as shown in Figs. S2b-e. The segregation coefficient of W and Re (only CMSX-4 contains Re) is much greater than 1, indicating that these heavy elements with high melting point enrich the dendrites, while the light elements with the low melting point, such as Al and Ti, partition to inter-dendrite. For dendrites solidified first during the solidification process, the addition of W and Re aggravates density inversion, thus enhancing the solute convection and fostering the freckle formation, which is consistent with the experimental results [16].

3.2 Relationship between crystal orientation and strain

Hurricanes can cause the branches to bend or even fracture, and similarly, researchers observed the sway and fracture of the dendrites due to ultrasound [44]. Similarly, dendrites may also deform and even fracture due to solute convection. From this point of view, there should be different states during the whole convection process as to strain or orientation. For example, strain accumulates during deformation, and is relieved to some extent after fracture/breakage. Therefore, we studied in the following the microstructure of freckles in the four brands of alloys, and analyzed the correlation between the orientation and strain of freckle grains to further understand the formation mechanism of freckles.

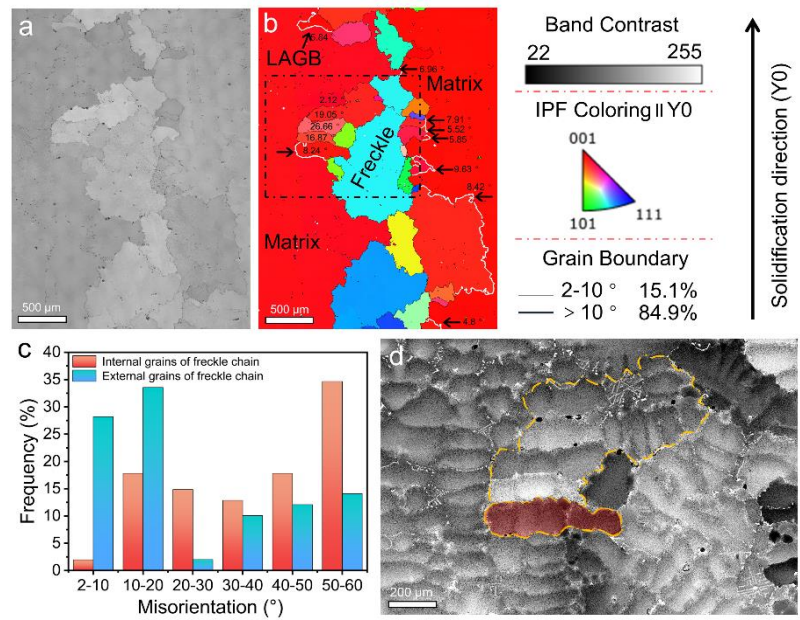

**Fig. 2.** MA distribution in freckle grains.

(a)-(b) Band contrast map and orientation map. LAGBs are highlighted with white lines and their MA are marked. (c) MA distribution for the internal and external grains.

(d) BSE image corresponding to the black square in (b), and the five freckle grains outlined by a yellow dashed line. Especially, the freckle grain marked in red will be studied for further HR-EBSD analysis.

Band contrast (BC) map not only reflects the quality of the Kikuchi pattern during EBSD data acquisition but also contains channel contrast, which can be used to indicate the grain orientation to some extent, so in combination with the BC map (Fig. 2a) and the orientation map (Fig. 2b), the area in the middle is determined to be freckles. It can be found from Fig. 2b that the freckle grains with smaller MA relative to the matrix mainly distribute near the interface between the freckle chain and matrix, while those with larger MA are mostly inside the freckle chain, in agreement with the statistics shown in Fig. 2c. Here, we define the MA of 2-10 ° as the low angle grain boundary (LAGB), which in Fig. 2b is marked with white curves and black arrows. Moreover, it is worth noting that in the BSE image (Fig. 2d) corresponding to the area denoted by a black square in Fig. 2b, five grains (outlined by a yellow dashed line) at the edge of freckle chains still maintain the secondary dendrite morphology, and are arranged on the primary dendrite, but they have small MA, meaning that they should be treated as freckle grains. Freckles in GTD111-DS, René N4 and CMSX-4 alloys were characterized, and similar results were obtained as shown in Supplementary Fig. S3. Therefore, it can be concluded that this phenomenon, freckle grains with smaller MA distribute near the interface and those with larger MA inside the freckle chain, is ubiquitous for Nickel-based superalloys with different composition and casting process.

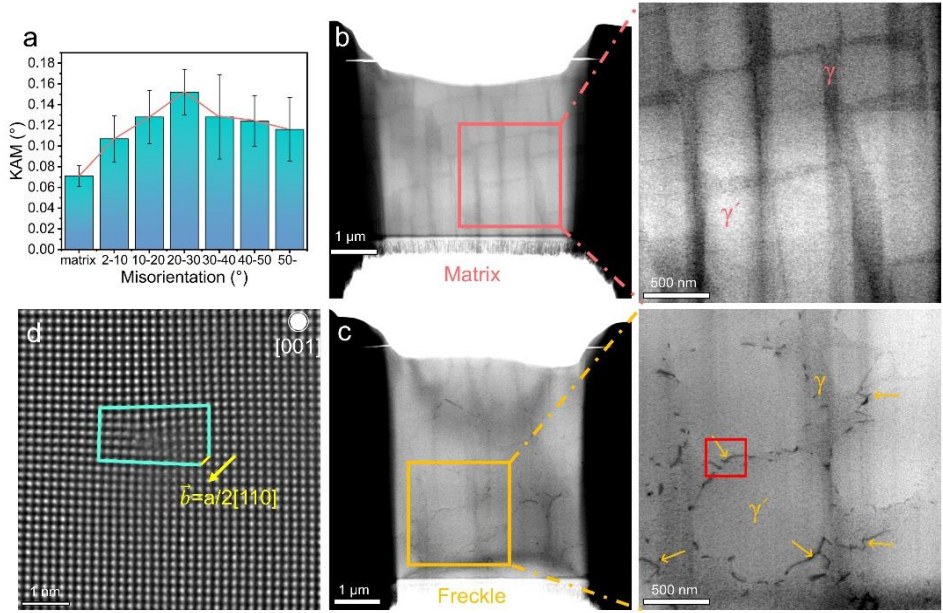

**Fig. 3.** Plastic deformation of matrix and freckle grains.

(a) Statistics of KAM variation with MA in four brands of alloys. The orange curve is used to describe the trend. (b) BF-STEM images of matrix. (c) BF-STEM images of freckle grain, and dislocations marked with yellow arrows. (d) High magnification HAADF-STEM image of dislocation core corresponding to the red square in (c), and Burgers vector determined by the Burgers circuit.

Kernel average misorientation (KAM) is usually used to characterize local strain qualitatively as it represents the average value of the misorientation between the pixel at the center of the kernel and every other pixel within the kernel, which essentially is the result of deformation [45]. Therefore, we use KAM here to analyze the strain of all 117 freckle grains in four alloys. To avoid the statistical error due to the limited sample number, we present the results every 10 degrees as shown in Fig. 3a. The abscissa represents the MA of freckle grains relative to the matrix, and the ordinate is the average KAM value of grains. We find that the KAM value, i.e., strain, increases at first, and reaches the maximum value in the range of 20-30 °, and then decreases with the further increase of the MA.

As we all know, once the external force on an object is removed, the change of shape and volume caused by elastic strain will be restored. However, the plastic strain will retain in the form of crystal defects such as dislocations. Under the convection of thermal-solute, the secondary dendrites are subjected to bending stress, which leads to the accumulation of plastic deformation. But with the dendrites fractured and even entering the mushy zone, it is no longer subjected to bending stress but only the thermal recovery of thermal-solute, which leads to the partial recovery of plastic deformation [46, 47]. Therefore, we suppose from Fig. 3a that the plastic deformation accumulated during the freckle formation is partially retained in the freckle grains, which is why the grains with MA above 50 °KAM value is still larger than the matrix. If it is true, there should be more defects such as dislocations or stacking faults in freckles than in the matrix.

To prove the above speculation, we made the TEM sample from the matrix and freckles, respectively. As shown in Figs. 3b-d, the matrix and freckles are all composed of γ and γ´ phases, but there are no obvious dislocations in the matrix (Figs. 3b), while lots of dislocations are found to grow along the γ/γ′ interface in freckles, marked with yellow arrows in the Fig. 3c, and one dislocation core is shown in Fig. 3d. According to Fig. 3, we can say that in the process of freckle formation, the secondary dendrite undergoes plastic deformation as a result of solute convection, giving rise to a number of dislocations, which gradually builds up enough crystallographic rotation. After the strain accumulates to a certain extent, grains fracture and then enter the mushy zone to rotate freely, the plastic strain partially retained in freckles. One more thing that should be noted is that in the above-mentioned discussion, freckles grains with large MA are formed generally earlier than the ones with small MA in the process of freckle formation as shown in Figs. 2-3, so grains formed earlier lay in the center of the freckle whereas those formed later along the interface.

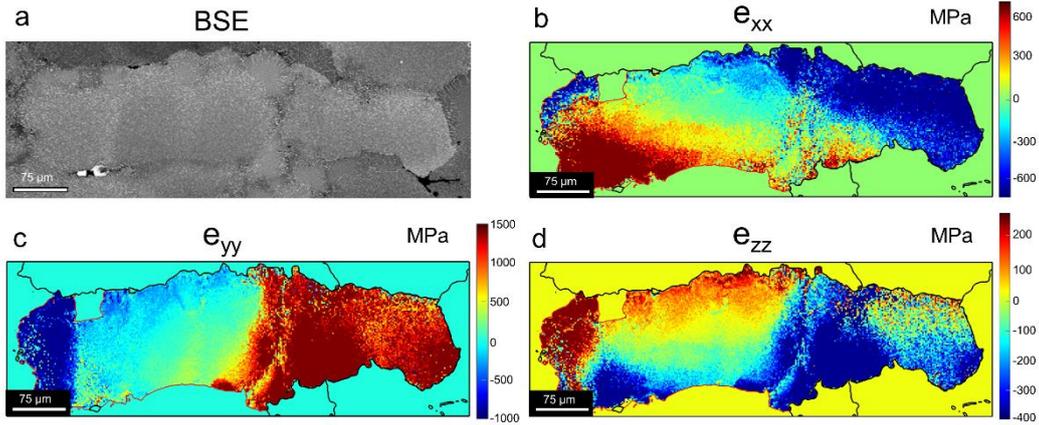

**Fig. 4.** Normal stress distribution obtained by HR-EBSD.
(a) BSE image of the freckle grain marked in red in Fig. 2d. (b)-(d) Stress distribution of $e_{xx}$, $e_{yy}$, and $e_{zz}$, respectively.

In order to further study the stress of dendrite caused by solute convection, we used high-resolution (HR) EBSD (with higher angular resolution than traditional EBSD) KAM to characterize the stress distribution inside the freckle grain. The grain in red with the MA 8.24 ° in Fig. 2d was characterized by HR-EBSD, and the quantitative stress distribution in the grain is shown in Fig. 4. We can see that the stress directions on both sides of the grain are always opposite, and the stress value in $e_{yy}$ (the direction of solute convection during directional solidification) is much larger. Meanwhile, the shear stress distributions have the same feature as shown in Supplementary Fig. S4. This can be explained that the root of the secondary dendrite is connected with the matrix, and the other side of the dendrite is under the solute convection, which makes the two sides of the secondary dendrite subjected to the opposite stress. Therefore, Fig. 4 describes the phenomena of strain distribution of freckles, which seems reasonable under the theory of solute convection.

Before the secondary dendrites fracture, strain gradually increases at the grain boundary nearby through deformation. In order to study the microstructure at the grain boundary, we combined EBSD and FIB techniques to accurately locate the interface and cut out TEM samples with a specific orientation, and performed the characterization by aberration-corrected electron microscope. As shown in the orientation map (Fig. S5a), we prepared the TEM sample with FIB at the position denoted by an arrow. Fig. S5b is the BF-STEM image of the interface, in which the left side is the matrix, the right side freckles, and the misorientation of LAGB in the middle is 4.64 °. The atomic resolution HAADF image of the interface (see Fig. S5c) shows that the two sides of the grain boundary are basically coherent when the matrix is in the orientation of [110]. To show the strain distribution at the interface geometric phase analysis (GPA) was carried out as shown in Fig. S5d, and a chain of a strain-concentrated area composed of a pair of compressive and tensile strain can be observed along the grain boundary. It is a dislocation core as shown in the high magnification HAADF image (Fig. S5e), and its Burgers vector determined by the Burgers circuit is $\vec{b}=a/2[110]$, which is common in this alloy. Therefore, the strain at the LAGB originates from a series of edge dislocations.

3.3 Deflection behavior of borides at the interface

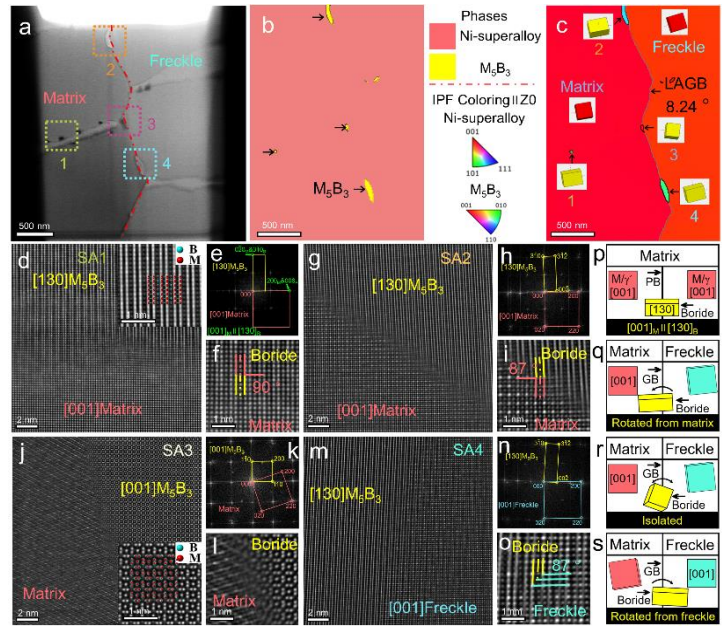

**Fig. 5.** Deflection behavior of borides.

(a) BF-STEM image of $M_5B_3$ borides sampled from the interface between the matrix and the freckle. And four borides were selected for further study, in which boride 1 locates at the γ/γ′ phase boundary, the others at the grain boundary. (b) TKD - phase distribution map, and borides marked with black arrows. (c) TKD - orientation map, and three-dimensional unit cells of matrix, freckles and borides showed to understand their orientation relationship. (d)-(o) HAADF-STEM, FFT and high magnification HAADF-STEM images of interface between alloys and borides1-4. (p)-(s) Schematic diagram of four borides, respectively.

As a grain boundary-strengthening element, element B is often added to Nickel-based superalloys to improve its creep properties [48, 49]. Usually, borides locate on the grain boundary [50], and there is a specific orientation relationship between borides and alloys [36-38]. Therefore, we can use the orientation relationship between borides and alloys to indirectly study the deflection behavior of freckle grains. A TEM sample was prepared at the grain boundary between the freckle grain marked in red and the left matrix in Fig. 2d, and the orientation and interfacial structure were characterized by TKD and aberration-corrected electron microscope as shown in Fig. 5. In the BF image shown in Fig. 5a, a LAGB is marked by a red dashed line, with the matrix on the left and the freckle on the right. Combined with TKD - phase distribution map (Fig. 5b), it can be seen that $M_5B_3$ borides exist at the γ/γ´ phase boundary (PB) and grain boundary (GB). EDS results (Fig. S6a) show that the heavy elements in borides are Cr and W, but for EDS is insensitive to B with small atomic number, EELS (Fig. S6b) was carried out and the signal of B is confirmed. By the combination of EDS and EELS results, the composition of borides is confirmed to be close to $M_5B_3$ shown in Fig. S6c, which is consistent with the calibration results of TKD. In order to understand the orientation relationship between borides and matrix or freckles more intuitively, three-dimensional unit cells of the matrix, freckle and Boride are overlayed on the IPF map in Fig. 5c.

We select four borides in Fig. 5a for further analysis. Boride 1 is located on the γ/γ´ interface in the matrix, and the HAADF image of the interface is shown in Fig. 5d, and the high magnification HAADF image of $M_5B_3$ in the orientation of [130] is inset top right. Because the contrast of the HAADF image is proportional to atomic number $Z^{1.7}$, it is impossible to image B atom with HAADF, so bright dots in the image represent M (here M denotes Cr and W) atoms, and the experimental image matches the model very well. Combined with the fast Fourier transformation (FFT) of the corresponding area (Fig. 5e), it can be seen that there is an orientation relationship of $[001]_M // [130]_B$ between borides and matrix, which is consistent with the literature [38]. As shown in the HAADF image of the interface, Fig. 5f, the matrix smoothly transits into the $M_5B_3$ phase and the interface is completely coherent, in agreement with the results of Fig. 5e.

Unlike boride 1, borides 2-4 are located at the LAGB between the matrix and freckles, and the deflection occurred in different degrees. The orientation relationship of boride 2 to the matrix is similar to boride 1, that is, borides are almost coherent with the matrix, which is confirmed especially by Figs. 5g-i. Boride 4 is almost the same, but it is coherent with the freckle grain but not the matrix (Figs. 5m-o). Borides 2 and 4 are deflected 3 ° in different directions on the basis of the original specific orientation. While for boride 3, there is a large misorientation as shown in Figs 5j-l.

According to Fig. 5, we know that borides in the matrix have a specific orientation relationship with the matrix, like boride 1, and the schematic diagram is shown in Fig. 5p. However, when borides locate at the interface between matrix and the freckles, there is MA between borides with the matrix or freckles, which should be due to the solute scouring. Three situations may exist for borides: almost attached to the matrix (Fig. 5q); almost attached to the freckle (Fig. 5s); and independent of matrix and freckles (Fig. 5r), just as borides 2, 4, and 3, respectively. In a word, we propose that solute scouring weakens the original orientation relationship between borides and alloys in different degrees, and therefore, the deflection behavior of borides relative to alloys can be understood based on the solute scouring on the dendrite in the solute convection theory.

## 4. Discussion

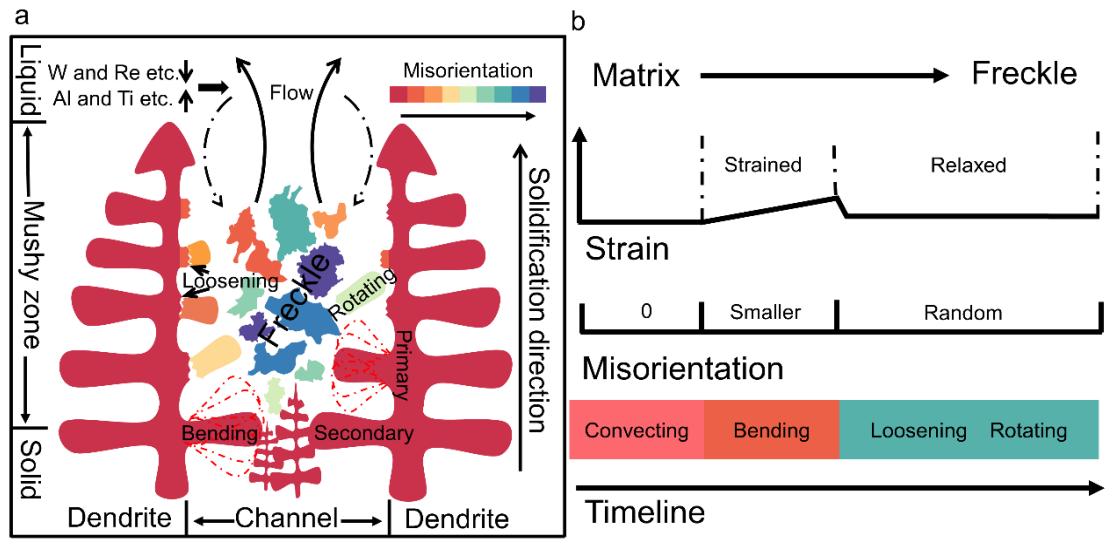

**Fig. 6.** Schematic diagram of freckle formation mechanism.

As shown in Fig. 6a, in the process of directional solidification, the alloy grows from bottom to top in the form of dendrites, with the primary dendrite orientation along the

< 001 > direction which is basically consistent with the heat flow direction. Generally, the alloy is divided into three areas from bottom to top: solid, mushy zone (mixture of solid and liquid) and liquid. In the horizontal direction, it is divided into two parts: a freckle channel in the middle and primary dendrites on two sides.

In the process of directional solidification, the dendrite solidifies first at the bottom, meaning heavy elements such as W and Re segregate. As a result, the density of the liquid at the bottom of the dendrite is lower than that near the tip of the dendrite on the top, and then the density inversion makes the liquid with low density in the mushy zone flow upward, thus solute convection is formed.

Due to the solute convection, the secondary dendrites on both sides of the channel will be scoured so that dendrites sway, loosen or even fracture. Thus, according to the different conditions of dendrites, freckle formation can be divided into two states, the strained state and the relaxed state, as shown in Fig. 6b. Local strain in the grains gradually increases with more deformation before fracture, which corresponds to the left side of the KAM peak (20-30 ° in Fig. 3a), meaning the strained state; while the relaxed state indicates that secondary dendrites are fractured and even enter the mushy zone, then the local strain in the grain decreases (the plastic strain is partially retained), which corresponds to the right side of the KAM peak. From the point of view of the MA, freckle grains are bended in the strained state due to the continuous scouring of solute, so the MA increases with more deformation. And when a limit is reached, secondary dendrites will loosen, it reaches the relaxed state, then the MA is larger than that in the strained state. When grains sink into the mushy zone, it rolls unrestrained and leads to the random orientation of freckle. Then when this area is solidified, freckle grains take shape, and with the growth of alloys along the <001> direction, more grains come to the channel, and freckle chains are formed in the end. Therefore, based on the results of the composition, strain and orientation analysis of Figs. 2-6, we speculate that the formation of freckles accords with the theory of solute convection including solute segregation, density inversion and dendrite fracture. But in fact, in the whole solidification process, the two most important influencing factors are the temperature field and solute field, while different solidification technology, alloy composition, casting structure and size will lead to the variation of these two factors, so it will be more complicated in the real solidification process. Due to the limitation of technical means, this paper only studies the microstructure after the formation of freckles, so still a lot of experimental and numerical simulation work needs to be done.

## 5. Conclusion

With the help of SEM, EBSD and aberration-corrected electron microscope, we study the composition, orientation and strain of freckle grains in four Nickel-based alloys from mesoscopic scale to microscopic scale. We find the different behavior of the internal stresses and the orientation from the freckle-matrix boundary to the freckle chain center. All experimental results give new proof to the theory of thermal-solutal convection and our finding as to the freckle formation is thought to be general, which can be applicable to both directionally solidified alloys and single crystal alloys, and is not affected by alloy composition and casting process parameters. Our results enrich the methodology of freckle studies, providing a new sight of the formation mechanism of casting defects.

## Declaration of interests

The authors declare no competing interests.

## Supplementary data

Supplemental Information can be found online.

## Acknowledgements

This work has been supported by Anhui Science and Technology Major Project (acceptance number: S2019b05050019), and is supported by the National Natural Science Foundation of China (No. 11874394), The University Synergy Innovation Program of Anhui Province (No. GXXT-2020-003).

## Author contribution

B.G supervised the project. S.W., Y.J., Y.Z., X.S., F.C. and B.G. conceived the experiments. Y.J. and Y.D. prepared turbine blades. S.W. and L.M. analyzed the EBSD data. Y.W. and Y.Z. performed the HREBSD experiments and analyzed the HREBSD data. S.W., Y.J., Y.W., Y.Z, L. M., Y.Z., X.S. and B.G prepared the manuscript. All authors discussed the results and contributed to the manuscript.